\documentclass[usletter, 12pt]{article}

\usepackage{amsmath,amsthm}
\usepackage{natbib}
\usepackage{geometry}
\usepackage{amssymb,authblk,amsfonts}
\usepackage{comment,enumerate,graphicx,hhline}

\usepackage[usenames,dvipsnames]{color, xcolor}

\usepackage{subcaption}
\usepackage[hidelinks]{hyperref}

\usepackage{amsmath,amsfonts,amstext,rotating,amssymb,graphics,xspace}

\usepackage{epsfig}
\usepackage{hyperref}
\hypersetup{
colorlinks   = true, 
urlcolor     = blue, 
linkcolor    = blue, 
citecolor   = blue    
}

\usepackage{amsmath}
\usepackage{natbib}
\usepackage{hyperref}
\usepackage{geometry}
\usepackage{type1cm}
\usepackage{mathrsfs}
\usepackage{amssymb}
\usepackage{authblk}
\usepackage{amsfonts}
\usepackage[T2A]{fontenc}
\usepackage{hhline}
\usepackage{amsthm}
\usepackage{multirow}
\usepackage{multicol}
\usepackage{color}
\usepackage{sgame}
\usepackage{mathtools}
\usepackage{caption}
\usepackage{graphicx}
\usepackage{wrapfig}
\usepackage{changepage}
\usepackage{lipsum}
\usepackage[english]{babel}
\usepackage[utf8]{inputenc}
\usepackage{comment}
\usepackage{tikz}
\usepackage{pgf}
\usetikzlibrary{positioning}
\usepackage{tabularx}
\usepackage{pbox}
\usepackage{subcaption}
\usepackage{enumitem}
\usepackage{rotating}
\usepackage{tcolorbox}
\usepackage{setspace}
\usepackage{array, makecell}

\usetikzlibrary{arrows,backgrounds}

\tikzstyle{object}=[circle,draw=red]
\tikzstyle{agent}=[circle,draw=blue]
\tikzstyle{quantity}=[fill=white]

\def\w{\omega}

\hypersetup{colorlinks = true}
\hypersetup{allcolors=blue}

\newtheorem{definition}{Definition}

\newtheorem{theorem}{Theorem}

\newcommand{\hide}[1]{} 

\usepackage{setspace}
\begin{document}
	
	\title{The core in the housing market model with fractional endowments
	}

	\author{Jingsheng Yu\thanks{School of Economics and Management, Wuhan University. Email: yujingsheng1987@outlook.com} \quad \quad Jun Zhang\thanks{Institute for Social and Economic Research, Nanjing Audit University. Email: zhangjun404@gmail.com}
	}
	
	\date{\today}

\maketitle	
	
\begin{abstract}\label{abstract}
      We explore the core concept in a generalization of the housing market model where agents own fractional endowments while maintaining ordinal preferences. Recognizing that individuals are easier than coalitions to block an allocation, we adopt a definition in which individuals block an allocation if their received assignments do not first-order stochastically dominate their endowment, while a non-singleton coalition blocks an allocation if they can reallocate their endowments to obtain new assignments that first-order stochastically dominate their original assignments. Our findings show that, unlike the original model, the strong core may be empty, while the weak core is nonempty. The weak core always contains elements that satisfy equal treatment of equals, but it may not contain elements satisfying equal-endowment no envy.
\end{abstract}	

\bigskip



\thispagestyle{empty}
\setcounter{page}{0}


\newpage

\section{Introduction}\label{section:intro}

This paper studies a generalization of the housing market model introduced by \cite{ShapleyScarf1974}, a foundational model in market design. In the original model, a finite set of agents each own distinct indivisible objects and seek to exchange them based on ordinal preferences. \cite{ShapleyScarf1974} and \cite{roth1977weak} approach this model from a cooperative game perspective. The weak core consists of allocations where no coalition of agents can strictly improve the welfare of every member by reallocating their endowments among themselves, while the strong core consists of allocations where no coalition can strictly improve the welfare of at least one member without making others worse off. 
When agents' preferences are strict, the strong core is nonempty and contains a unique element, which coincides with the outcome of the top trading cycle mechanism. This paper extends the model to allow agents to own fractional amounts of objects while maintaining ordinal preferences. This setup naturally arises when agents hold partial ownership and co-own objects with others. We examine The strong core in this context, following the approaches of \cite{ShapleyScarf1974} and \cite{roth1977weak}.

A consequence of this generalization is the need to extend agents' ordinal preferences to assignments they may receive in any allocation. The extension based on first-order stochastic dominance used in the literature is an incomplete relation. As a result, there are multiple ways to define coalition blocking.\footnote{This multiplicity also arises in the definition of incentive properties in the random assignment literature, such as the distinction between strategy-proofness and weak strategy-proofness \citep{bogomolnaia2001new}, as well as between group manipulation and strong group manipulation \citep{bade2016random,zhang2019efficient}.} One possible definition states that a coalition will block an allocation if they can reallocate their endowments to obtain new assignments that first-order stochastically dominate their current assignments. Another possibility says that a coalition will block an allocation if they can reallocate their endowments to obtain new assignments that are not first-order stochastically dominated by their current assignments. The former definition imposes a more stringent requirement, leading to a larger core.

This paper adopts a definition between the two possibilities outlined above. We believe that individuals are easier than coalitions of two or more agents to block an allocation, as the latter requires coordination. Therefore, we assume that an agent will block an allocation if their received assignment does not first-order stochastically dominate their endowment, while a non-singleton coalition will block an allocation if they can reallocate their endowments to obtain new assignments that first-order stochastically dominate their current assignments. This individual blocking requirement is equivalent to the individual rationality condition commonly used in the literature. We then define the strong core and weak core similarly to the original model. Our findings show that, assuming strict preferences, the strong core may be empty, while the weak core is always nonempty but compatible only with mild fairness criteria. In every economy, there exists a weak core allocation that satisfies equal treatment of equals (ETE), but in some economies, all weak core allocations violate equal-endowment no envy (EENE). ETE is a mild fairness criterion, requiring that two agents with identical preferences and endowments receive identical assignments. EENE is stronger and intuitive, requiring that two agents with equal endowments prefer their own assignment over the other's. Since we are considering an endowment exchange model, if two agents contribute the same resources to the economy, any differences in their final assignments should be attributed solely to their preferences.

\cite{AS2011} study a more general setup than ours. They propose a mechanism satisfying individual rationality, sd-efficiency, and no justified envy, but violating EENE. However, they support EENE as a desirable fairness criterion. \cite{YuzhangFTTCnew} propose a class of mechanisms that satisfy individual rationality, sd-efficiency, and EENE for our model. Thus, our findings imply that the outcomes of their mechanisms do not belong to the weak core defined in this paper.

Our proof of the nonemptiness of the weak core relies on a theorem of \cite{mas1992equilibrium} by constructing cardinal utilities for agents and utilizing the existence of a Walrasian equilibrium with slack. This approach is similar to the method used by \cite{basteck2018fair} to prove the nonemptiness of the core from equal division in the house allocation model, which is a special case of our model where agents own an equal division of all objects. However, our results are not simple extensions of \cite{basteck2018fair} due to the complex endowments in our model. For example, while the strong core from equal division is nonempty in the house allocation model, it can be empty in our model. It remains an open question whether an algorithm exists that uses only agents' ordinal preference information to find the elements of the weak core that satisfy ETE.

\section{Housing markets with fractional endowments}\label{section:model}

There are $ n $ agents $ I=\{1,2,\ldots,n\} $ and $ n $ objects $ O=\{o_1,o_2,\ldots,o_n\} $. Each $ i\in I $ is endowed with an amount $ \w_{i,o}\in [0,1] $ of each $ o\in O $ such that $ \sum_{i\in I}\w_{i,o}=1 $ and $ \sum_{o\in O}\w_{i,o}=1 $. When objects are indivisible, $ \w_{i,o} $ is the probability share of $ o $ owned by $ i $. We use a doubly stochastic matrix $ \w=(\w_{i,o})_{i\in I,o\in O} $ to denote the endowment distribution. Then, each row $ \w_i=(\w_{i,o})_{o\in O} $ of the matrix represents the endowments of $ i $. Each $ i $ has a strict preference relation $ \succ_i $ over objects. For all $ o, o'\in O $, we write $ o\succ_I o' $ if $ o\succ_i o' $ or $ o=o' $. The preference profile of agents is represented by $ \succ_I= (\succ_i)_{i\in I} $. An economy is represented by $ (I,O,\succ_I,\w) $.

In an economy $ (I,O,\succ_I, \w) $, an \textbf{assignment} for agent $ i $ is a vector $ p_i\in [0,1]^n $ such that $ \sum_{o\in O}p_{i,o} =1 $. Given preferences $ \succ_i $, $p_i$ \textbf{weakly (first-order) stochastically dominates} $p_i'$, denoted by $ p_i\succsim^{sd}_i p'_i $, if $ \sum_{o^{\prime }\succsim _{i}o}p_{i,o^{\prime }}\geq\sum_{o^{\prime }\succsim _{i}o}p'_{i,o^{\prime }}$ for all $ o\in O $. If the inequality is strict for some $ o $, $p_i$ \textbf{strictly stochastically dominates} $p'_i$, denoted by $ p_i \succ^{sd}_i p'_i $.

An \textbf{allocation} is represented by a matrix $p=(p_{i,o})_{i\in I, o\in O}\in [0,1]^{n \times n}$ such that, for all $ o\in O$, $\sum_{i\in I}p_{i,o}= 1$, and, for all $ i\in I$, $\sum_{o\in O}p_{i,o}=1$. For each $ i\in I $, $p_{i}=(p_{i,o})_{o\in O}$ is the assignment received by $ i $.  An allocation $p$ weakly stochastically dominates another $p^{\prime }$, denoted by $p \succsim^{sd}_I p^{\prime }  $, if $p_{i}\succsim^{sd}_i p_{i}^{\prime }$ for all $i\in I$; $ p $ strictly stochastically dominates $ p' $, denoted by $p \succ^{sd}_I p^{\prime }  $, if there further exists an agent $ j $ such that $p_{j} \succ^{sd}_j p_{j}^{\prime }$.

An allocation $ p $ is \textbf{sd-efficient} if it is not strictly stochastically dominated by any other allocation.  It is \textbf{individually rational} (IR) if, for all $ i\in I $, $ p_i \succsim^{sd}_i \w_i $.

In an allocation $ p $, agent $ i $ is said to \textbf{envy} another $ j $ if $ p_i\not\succsim^{sd}_i p_j $. An allocation $ p $ satisfies \textbf{equal treatment of equals} (ETE) if, for all $ i,j\in I $ with $ \w_i=\w_j $ and $ \succ_i=\succ_j $, $ p_i=p_j $. An allocation $ p $ satisfies \textbf{equal-endowment no envy} (EENE) if, for all $ i,j \in I$ with $ \w_i=\w_j $, $ p_i\succsim^{sd}_i p_j $ and $ p_j\succsim^{sd}_j p_i $. EENE implies ETE.

\section{Main result}\label{section:core}

For individuals, we impose a blocking condition equivalent to requiring IR: as long as an allocation $ p $ is not individually rational for an agent $ i $, $ i $ will block $ p $.

For coalitions of two or more agents, we impose a stricter requirement: to block an allocation, they must reallocate their endowments to obtain new assignments that stochastically dominate their original assignments.

\begin{definition}
	A non-singleton coalition $ I'\subseteq I $ \textbf{weakly block} an allocation $ p $ via another allocation $ p' $ if $ \sum_{i\in I'} p'_i=\sum_{i\in I'}\w_i $, for all $ i\in I' $, $ p'_i\succsim^{sd}_i p_i $,  and for some $ j\in I' $, $ p'_j \ \succ^{sd}_j \ p_j $. If $ p'_i \succ^{sd}_i p_i $ for all $ i\in I' $, then we say that $ I' $ \textbf{strongly block} $ p $ via $ p' $. 	
\end{definition}

\begin{definition}
	In any economy, the \textbf{strong core} consists of individually rational allocations that are not weakly blocked by non-singleton coalitions; the \textbf{weak core} consists of individually rational allocations that are not strongly blocked by non-singleton coalitions.
\end{definition}

The strong core is a subset of the weak core. In the housing market model, they reduce to the familiar definitions used by \cite{ShapleyScarf1974} and \cite{roth1977weak}.

Our main result is the following theorem.

\begin{theorem}\label{thm}
	In the housing markets with fractional endowments model:
	\begin{enumerate}
		\item The strong core may be empty.
		
		\item The weak core is nonempty, and always contains an element satisfying ETE.
		
		\item There exist economies in which every element of the weak core violates EENE.
	\end{enumerate}
\end{theorem}

\section{Proof of \autoref{thm}}\label{section:proof}

\subsection{Proof of the first and third statements}

We prove examples to prove the first and third statements.

Consider an economy with four agents $ I=\{1,2,3,4\} $ and four objects $ O=\{o_1,o_2,o_3,o_4\} $. Agents have the following endowments: $ 1 $ and $ 2 $ each own $ 1/2o_1 $ and $ 1/2o_3 $, and $ 3 $ and $ 4 $ each own $ 1/2o_2 $ and $ 1/2o_4 $.  We consider two preference profiles shown below, $ \succ_I $ and $ \succ'_I $.
	\begin{table}[!h]
		\centering
		\begin{subtable}{.3\linewidth}
			\centering
			\begin{tabular}[c]{c|cccc}
				& $o_{1}$ & $o_{2}$ & $o_{3}$ & $o_{4}$ \\\hline
				$1$ & $1/2$ &  & $1/2$ &  \\
				$2$ & $1/2$ &  & $1/2$ &  \\
				$3$ &  & $1/2$ &  & $ 1/2 $ \\
				$4$ &  & $1/2$ &  & $ 1/2 $
			\end{tabular}
			\subcaption{Endowments}
		\end{subtable}
		\quad
		\begin{subtable}{.3\linewidth}
			\centering
			\begin{tabular}[c]{cccc}%
				$\succ_{1}$ & $\succ_{2}$ & $\succ_{3}$ & $\succ_{4}$ \\\hline
				$o_2$ & $o_2$ & $o_1$ & $ o_2 $ \\
				$o_1$ & $o_1$ & $o_2$ & $ o_1 $ \\
				$o_4$ & $o_4$ & $o_3$ & $ o_4 $\\
				$ o_3 $ & $ o_3 $ & $ o_4 $ & $ o_3 $
			\end{tabular}
			\subcaption{$ \succ_I $}
		\end{subtable}
		\quad
		\begin{subtable}{.3\linewidth}
			\centering
			\begin{tabular}[c]{cccc}%
				$\succ_{1}$ & $\succ'_{2}$ & $\succ_{3}$ & $\succ'_{4}$ \\\hline
				$o_2$ & $o_1$ & $o_1$ & $ o_1 $ \\
				$o_1$ & $o_2$ & $o_2$ & $ o_2 $ \\
				$o_4$ & $o_4$ & $o_3$ & $ o_4 $\\
				$ o_3 $ & $ o_3 $ & $ o_4 $ & $ o_3 $
			\end{tabular}
			\subcaption{$ \succ'_I $}
		\end{subtable}
	\end{table}
	
	\textbf{The strong core is empty under $ \succ_I $}.
	
	 We prove that the strong core is empty when agents' preferences are $ \succ_I $. Consider any IR allocation $ p $. Because every $ i $ prefers $ o_1$ and $o_2 $ over $ o_3$ and $o_4 $, and $ \w_{i,o_1}+\w_{i,o_2} =\w_{i,o_3}+\w_{i,o_4}=1/2$, IR requires that $ p_{i,o_1}+p_{i,o_2}=1/2 $ and $ p_{i,o_3}+p_{i,o_4}=1/2 $. Moreover, since $ 4 $ prefers $ o_2 $ over $ o_1 $ and prefers $ o_4 $ over $ o_3 $, IR requires that 4 must receive his endowments $ (1/2o_2,1/2o_4) $.
	 
	  Given this, we notice that if $ 1 $ and $3 $ exchange endowments, they will respectively obtain the best assignments they are possible to obtain in any IR allocation. Symmetrically, if $ 2 $ and $3 $ exchange endowments, they will respectively obtain the best assignments they are possible to obtain in any IR allocation. So, $ p $ is not weakly blocked by $ \{1,3\} $ through exchanging endowments if and only if $ p_{1}=\w_{3} $ and $ p_{3}=\w_{1} $, and $ p $ is not weakly blocked by $ \{2,3\} $ through exchanging endowments if and only if $ p_{2}=\w_{3} $ and $ p_{3}=\w_{2} $. Therefore, $ p $ is in the strong core if and only if $ p_{1}=p_{2}=\w_{3} $, $ p_{3}=\w_{1}=\w_{2} $, and $ p_4=\w_4 $, which is impossible. This means that the strong core is empty in the economy with $ \succ_I $.

	\bigskip
	
	\textbf{The weak core under $ \succ'_I $ violates EENE}.
	
	 We prove that when agents' preferences are $ \succ'_I $, all elements of the weak core violates EENE. The nonemptiness of the weak core is proved in the next subsection.
	 
	  Let $ p $ be any IR allocation that satisfies EENE. Because every agent $ i $ prefers $ o_1$ and $o_2 $ over $ o_3$ and $o_4 $, similar to the above, IR requires that $ p_{i,o_1}+p_{i,o_2}=1/2 $ and $ p_{i,o_3}+p_{i,o_4}=1/2 $. Moreover, because $ 2 $ prefers $ o_1 $ over $ o_2 $ and $ 4 $ prefers $ o_4 $ over $ o_3 $, IR requires that $ 2 $ must receive his endowment $ 1/2o_1 $ and $ 4 $ must receive his endowment $ 1/2o_4 $. 
	  
	  Because $ 1 $ and $ 2 $ own identical endowments and both prefer $ o_4 $ over $ o_3 $, EENE requires that $ p_{1,o_3}=p_{2,o_3} $ and $ p_{1,o_4}=p_{2,o_4} $. Similarly, because $ 3 $ and $4 $ own identical endowments and both prefer $ o_1 $ over $ o_2 $, EENE requires that $ p_{3,o_1}=p_{4,o_1} $ and $ p_{3,o_2}=p_{4,o_2} $. So, given $ p_{2,o_1}=1/2 $ and $ p_{4,o_4} =1/2$, it must be that $ p_{1,o_4}=p_{2,o_4}\le 1/4 $ and $ p_{3,o_1}=p_{4,o_1}\le 1/4 $. These imply that $ p_{3,o_2}=p_{4,o_2}\ge 1/4 $.  Then,  $ p_{1,o_2}\le 1/2 $. Thus, we obtain $ p_{1,o_2}\le 1/2 $ and $ p_{1,o_4}\le 1/4 $, and $ p_{3,o_1}\le 1/4 $. This implies that $ \{1,3\} $ will strongly block $ p $ by exchanging  endowments such that $ 1 $ obtains $ (1/2o_2,1/2o_4) $ and $ 3 $ obtains $ (1/2o_1,1/2o_3) $. Therefore, $ p $ is not in the weak core.
	
\subsection{Proof of the second statement}	

Although agents' preferences are ordinal in our model, our proof endows agents with cardinal utilities that are consistent with their ordinal preferences. In particular, we endow agents of equal preferences and endowments with equal utilities. We prove the existence of a Walrasian equilibrium with slack by using the main theorem of \cite{mas1992equilibrium}. We then construct a sequence of Walrasian equilibria with slack and prove that their limit belongs to the weak core and satisfies ETE.

Formally, in an economy, we endow each agent $ i $ with von Neumann-Morgenstern utilities $ (u_{i,o})_{o\in O}\in \mathbf{R}_{+}^{n} $ such that $ o\succ_i o' \Longleftrightarrow u_{i,o}>u_{i,o'} $. If two agents have equal endowments and preferences, they are endowed with equal utilities. The utility of obtaining nothing is normalized to zero. We assume that agents compare assignments by expected utilities. Every $ i $'s expected utility from an assignment $ p_i $ is $ u_i(p_i)=\sum_{o\in O} u_{i,o}\cdot p_{i,o} $. 
	
For any $ \varepsilon>0 $, we define $ i $'s consumption space as
	\[
	X^\varepsilon_i=\{x_i \in [0,1]^n: \sum_{o\in O}x_{i,o}\le 1, \sum_{o'\succsim_i o}x_{i,o'}\ge \sum_{o'\succsim_i o}\w_{i,o'}-\varepsilon \text{ for all }o\in O\}.
	\]
So $ X^\varepsilon_i $ is the set of assignments that are close to be IR for $ i $ with an error bounded by $ \varepsilon $.	It is clear that $ X^\varepsilon_i $ is nonempty, closed, and convex.

We then define the space of price vectors to be 
	\[
	\Delta=\{P\in \mathbf{R}^{n}_+:\sum_{o\in O}P_o\le 1\}.
	\]
	
 It is easy to verify that the conditions (I)--(VI) in \cite{mas1992equilibrium} are satisfied.\footnote{\cite{mas1992equilibrium} study a general equilibrium model with consumers and firms. We refer the reader to the paper for details.} His condition (V) is satisfied because $ \varepsilon>0 $ ensures that, for every nonzero $ P\in  \Delta $, $ P\cdot \w_i>\inf_{x_i\in X^\varepsilon_i} P\cdot x_i $. By \citeauthor{mas1992equilibrium}'s Theorem 1, for any $ \varepsilon>0 $, there exists a \textit{Walrasian equilibrium with slack} (WE-slack) $ (x^\varepsilon,P^\varepsilon,\alpha^\varepsilon) $ such that
	\begin{enumerate}
		\item  $ P^\varepsilon\in \Delta $ and $ \alpha^\varepsilon \ge 0 $;
		
		\item $ \sum_{i\in I}x^\varepsilon_{i,o}=\sum_{i\in I}\w_{i,o} $ for all $ o\in O $;
		
		\item $ x^\varepsilon_i\in\arg\max\{u_i(x_i):x_i\in X^\varepsilon_i,P^\varepsilon\cdot x_i \le P^\varepsilon\cdot \w_i+\alpha^\varepsilon \} $ for all $ i\in I $.
	\end{enumerate} 
	The facts that $ x^\varepsilon_i\in X^\varepsilon_i $ for all $ i\in I $ and that $ \sum_{i\in I}x^\varepsilon_{i,o}=\sum_{i\in I}\w_{i,o} $ for all $ o\in O $ imply that $ x^\varepsilon $ is an allocation. Actually, $ x^\varepsilon $ can be chosen to satisfy ETE. To see this, let $ I_1,\cdots, I_K $ denote the partition of agents such that each $ I_\ell $ is a subset of agents of equal preferences and endowments. Given any WE-slack $ (x^\varepsilon,P^\varepsilon,\alpha^\varepsilon) $, for every $ I_\ell $ and every $ i\in I_\ell $, define
	\[
	y^\varepsilon_i=\dfrac{\sum_{j\in I_\ell}x^\varepsilon_j}{|I_\ell|}.
	\]
	It is clear that $ y^\varepsilon_i\in X^\varepsilon_i $, $ P^\varepsilon \cdot y^\varepsilon_i\le  P\cdot \w_i+\alpha^\varepsilon$, and $ y^\varepsilon_i $ solves $ i $'s expected utility maximization problem (since expected utility functions are linear). Also, $ \sum_{i\in I}y^\varepsilon_{i,o}=\sum_{i\in I}\w_{i,o} $ for all $ o\in O $. So, $ (y^\epsilon,P^\varepsilon,\alpha^\varepsilon) $ is a WE-slack and $ y^\epsilon $ satisfies ETE.
	
	Let $ (\varepsilon^k)_{k\in \mathbb{N}} $ be a sequence of positive real numbers such that $ \varepsilon^k\rightarrow 0 $ as $ k\rightarrow \infty $. For each $ \varepsilon^k $, let $ (x^k,P^k,\alpha^k) $ be a WE-slack that satisfies ETE. Since $ (x^k)_{k\in \mathbb{N}} $ is bounded, it has limit points (the limits of convergent subsequences). Let $ x^* $ be any limit point. Since each $ x^k $ is an allocation satisfying ETE, $ x^* $ must be an allocation satisfying ETE. Moreover, for every agent $ i $, $ x^k_i\in  X^{\varepsilon^k}_i $ implies that $ x^*_i\in X^0_i $, which means that $ x^* $ is individually rational.
	
	We prove that $ x^* $ belongs to the weak core. Suppose towards a contradiction that $ x^* $ is strongly blocked by a non-singleton coalition $ I' $ via another allocation $ x' $. Then, for every $ i\in I' $, $ u_i(x'_i)>u_i(x^*_i) $. Since $ x^* $ is a limit point of $ (x^k)_{k\in \mathbb{N}} $, there exists a large enough $ n\in \mathbb{N} $ such that, for all $ i\in I' $, $ u_i(x'_i)>u_i(x^n_i) $. We also have that, for all $ i\in I' $, $ x'_i \in X^{\varepsilon^n}_i $. So, for all $ i\in I' $, it must be that $ P^n\cdot x'_i> P^n\cdot \w_i+\alpha^n\ge P^n\cdot \w_i$. This implies that $ P^n\cdot \sum_{i\in I'}x'_i>P^n\cdot \sum_{i\in I'}\w_i$, which contradicts that $ \sum_{i\in I'}x'_i = \sum_{i\in I'}\w_i$.
	
	This finishes the proof of the second statement.

\bibliographystyle{aer}

\begin{center}
	\begingroup
	\setstretch{1.0}
	\bibliography{reference}
	\endgroup
\end{center}

\end{document}